\documentclass[twocolumn,showpacs,preprintnumbers,amsmath,amssymb,prb]{revtex4}

\usepackage{graphicx}
\usepackage{bm}
\usepackage{subfigure}

\begin{document}

\title{
\bf\Large{Crossover from $\bm{hc/e}$ to $\bm{hc/2e}$
current oscillations 
\\ in rings of $\bm s$-wave superconductors}}

\author{F. Loder$^1$}
\author{A. P. Kampf$^1$}
\author{T. Kopp$^1$}
\affiliation{
$^1$Center for Electronic Correlations and Magnetism, Institute of Physics, University of 
Augsburg, D-86135 Augsburg, Germany}

\date{\today}

\begin{abstract}
We analyze the crossover from an $hc/e$-periodicity of the persistent current in
flux threaded 
clean metallic rings towards an $hc/2e$-flux periodicity of the 
supercurrent upon entering the superconducting state. On the basis of a model 
calculation for a one-dimensional ring we identify the 
underlying mechanism, which balances the $hc/e$ versus the $hc/2e$ periodic components 
of the current density.
When the ring circumference exceeds the coherence length of 
the superconductor, the flux dependence is strictly $hc/2e$ periodic.
Further, we develop a multi-channel model which reduces the Bogoliubov - de Gennes equations to
a one-dimensional differential equation for the radial component of the wave function. The discretization
of this differential equation introduces transverse channels, whose number scales with the thickness
of the ring. The periodicity crossover is analyzed close the critical temperature. 
\end{abstract}


\maketitle

\section{Introduction}

Charged particles, which encircle a magnetic flux threaded region, acquire a 
geometric phase. This Aharonov-Bohm (AB) phase leads to quantum
interference phenomena along multiply connected paths \cite{AB}. A particular 
manifestation of the AB-effect is the persistent current in mesoscopic metal
rings~\cite{Landauer, landauer:85}, which is modulated periodically by the magnetic flux piercing the 
interior of the ring with the period of a flux quantum $\Phi_0=hc/e$ for clean rings. 

Likewise, in superconducting rings the order parameter responds periodically 
to magnetic flux, as implied by the requirement of a single-valued 
superconducting wave function in the presence of a supercurrent 
\cite{London,Byers,schrieffer}. Measurements of magnetic flux trapped in small
cylinders proved that the flux in superconductors is quantized in units of 
$\Phi_0/2$ \cite{Doll,Deaver}. The $hc/2e$ superconducting flux 
quantum was corroborated by measurements of the $hc/2e$ periodicity of the 
critical temperature of superconducting rings by Little and Parks 
\cite{Little,Parks}, and by the $hc/2e$ flux quantization of Abrikosov vortices 
\cite{Essmann}. 

The oscillations of the persistent current or the supercurrent with respect to
the magnetic flux implies the corresponding periodicity for all thermodynamic 
functions \cite{degennes}.
Two classes of condensate states have been identified which are not related by
a gauge transformation. In  the thermodynamic limit, they are degenerate  for integer and half integer flux values, which results in the observed $\Phi_0/2$ periodicity. 
This degeneracy is however lifted for discrete systems \cite{kulik:75}, which was
implicitly understood already in the early works of Byers and Yang \cite{Byers} and by Brenig \cite{brenig:61}.
The lifting of the degeneracy can be made explicit through the evaluation of the supercurrent in sufficiently small rings \cite{loder:07}. Recently, nodal superconductors have been in the focus of research \cite{loder:07,barash:08, tesanovic:08,zhu:08} as they allow for striking differences in the excitation spectrum for flux sectors centered  around integer and half integer $\Phi_0$ values, respectively.

For rings of $s$-wave superconductors, it is expected, that the $hc/e$ periodicity is restored if, the ring diameter
is smaller than the coherence length  \cite{czajka:05,loder:07,goldbart:08,vakaryuk:08}.
While the $\Phi_0$ and the $\Phi_0/2$ periods are well understood in
metallic and  superconducting rings, it has remained
unaddressed how the periodicity evolves for such a small ring when the normal metal turns
superconducting.
Here we analyze the periodicity crossover in a one-dimensional (1$D$) 
model for a flux threaded ring at zero temperature, which allows for a transparent analytical 
treatment on the basis of the Gor'kov equations in an external magnetic field
(Sec.~\ref{sec:1D}).

The gap equation for the current carrying superconducting state is solved for 
finite size rings to evaluate the field dependence of the discrete energy 
spectrum and the supercurrent. 
We identify two components of the current with 
$hc/e$ and $hc/2e$ periodicity, respectively, whose magnitudes shift with 
the opening
and increase of the energy gap in the superconducting 
state. When the coherence length of the superconducting ring is of the order 
of the ring size, only the $hc/2e$-periodic component remains. 
A similar analysis for the temperature
driven crossover in clean and dirty 1$D$ rings has recently been published by Wei and Goldbart \cite{goldbart:08}.

It is well known that a long-range ordered superconducting state does not exist in 1$D$. 
However, whereas thermal phase slips suppress a transition into the superconducting state 
at finite temperature, quantum phase slips at zero temperature are rare events. Even if phase coherence is broken at certain instants in time and 
space, the supercurrent does not decay in the ring. We therefore investigate only the zero temperature transition for the 1$D$ ring.

Subsequently (Sec.~\ref{sec:annulus}) we extend our analysis 
to rings of finite thickness (``annuli''),
which represent, from a formal point of view, multichannel systems. For the annuli we present the periodicity crossover
upon cooling through the superconducting transition temperature. For annular systems, that are confined to
a 2$D$ plane, we introduce a semi-analytical approach in which the numerical work is
reduced to the solution of a one-dimensional differential equation for the radial component
of the wave function. Its discretization allows to introduce a fixed number of (transverse) channels, whose number parameterizes the thickness of the annulus. 

\section{1D Ring}\label{sec:1D}

We start from the tight binding form of the kinetic energy for a 1$D$ ring 
with $N$ sites as given by
\begin{equation}
{\cal H}_0=-t\sum_{\langle ij\rangle,s}e^{\varphi_{ij}}c_{js}^\dag c_{is},
\label{s01}
\end{equation}
where the sum extends over all nearest-neighbor sites $i$ and $j$; $s=\uparrow,
\downarrow$ denotes the spin, and $t$ is the hopping matrix element. The 
vector potential $\bf A$ of an external magnetic field enters through the 
Peierls phase factor $\varphi_{ij}=(e/\hbar c)\int_i^j\text{\bf A}\cdot d{\bf r}=2
\pi\phi/N$, where $\phi=\Phi/\Phi_0$ and $\Phi$ is the magnetic flux through 
the ring. After Fourier transformation ${\cal H}_0$ becomes
\begin{equation}
{\cal H}_0=\sum_{ks}\epsilon_{k-\phi}c_{ks}^\dag c_{ks},
\label{s04}
\end{equation}
with the single-particle energy for a state with angular momentum $\hbar k$
\begin{equation}
\epsilon_{k-\phi}=-2t\cos\left(\frac{k-\phi}{R}\right).
\label{s05}
\end{equation}
$R=N/2\pi$ denotes the dimensionless radius of the ring and $k=-N/2,\dots,N/2-1$. If 
$N$ is a multiple of 4, a $k$-value exists for which $\epsilon_k=0$ 
\cite{cheung:88}, 
with two states exactly at the Fermi energy $E_F=0$ for $\phi=0$.
To ensure a unique ground state, we 
choose $\mu=t/N$ which is placed in between two single-particle 
energies $\epsilon_k$. This is achieved for even $N$, which are not multiples 
of 4. All calculations were performed for this generic choice of $N$ and $\mu$.

The superconducting state in this strictly 1$D$ ring model is 
controlled by a BCS-type Hamiltonian of the form
\begin{equation}
{\cal H}={\cal H}_0+\sum_{k,q}\Big[\Delta^*_k(q)c_{-k+q\downarrow}c_{k\uparrow}
+\Delta_k(q)c^\dag_{k\uparrow}c^\dag_{-k+q\downarrow}\Big],
\label{s09}
\end{equation}
where $\Delta_k(q)$ is the superconducting order parameter for the formation 
of Cooper pairs with finite angular momentum $\hbar q$ and $q\in\mathbb{Z}$. 
The order parameter is obtained from the anomalous imaginary time Green's 
function $F(k,k',\tau-\tau')=\langle T_\tau c_{k\downarrow}(\tau)c_{-k'
\uparrow}(\tau')\rangle$ \cite{mineev17} 
by
\begin{align}
\Delta_k(q)&=k_BT\sum_{k'}\sum_nV_{kk'}F(k',k'-q,\omega_n),\label{s011.1}
\end{align}
where $\omega_n=(2n-1)\pi k_BT$ is the fermionic Matsubara frequency for temperature 
$T$ and $V_{kk'}$ is the pairing interaction; $T_\tau$ is the time-ordering 
operator.

$\Delta_k(q)$ has to be determined self-consistently in the superconducting 
state. This is achieved by solving the equations of motion for the anomalous 
Green's function and the single particle propagator $G(k,\tau-\tau')=\langle 
T_\tau c_{ks}(\tau)c^\dag_{ks}(\tau')\rangle$, which is diagonal with respect 
to momentum and spin. This leads to the self-consistent set of Gor'kov 
equations:
\begin{multline}
G^{-1}(k,\omega_n)={G_0}^{-1}(k,\omega_n)\\+\sum_{q}\Delta_k(q)G_0(-k+q,-\omega_n)\Delta_{k-q}^*
(q),
\label{s4}
\end{multline}
\begin{align}
F(k,k-q,\omega_n)&={G_0}(k,\omega_n)\Delta_k(q)G(-k+q,-\omega_n),
\label{s6}
\end{align}
where $G_0(k,\omega_n)=[i\hbar\omega_n-\epsilon_{k-\phi}]^{-1}$ is the Green's function in the 
normal state.

\begin{figure}[t]
\centering
\includegraphics[width=84mm]{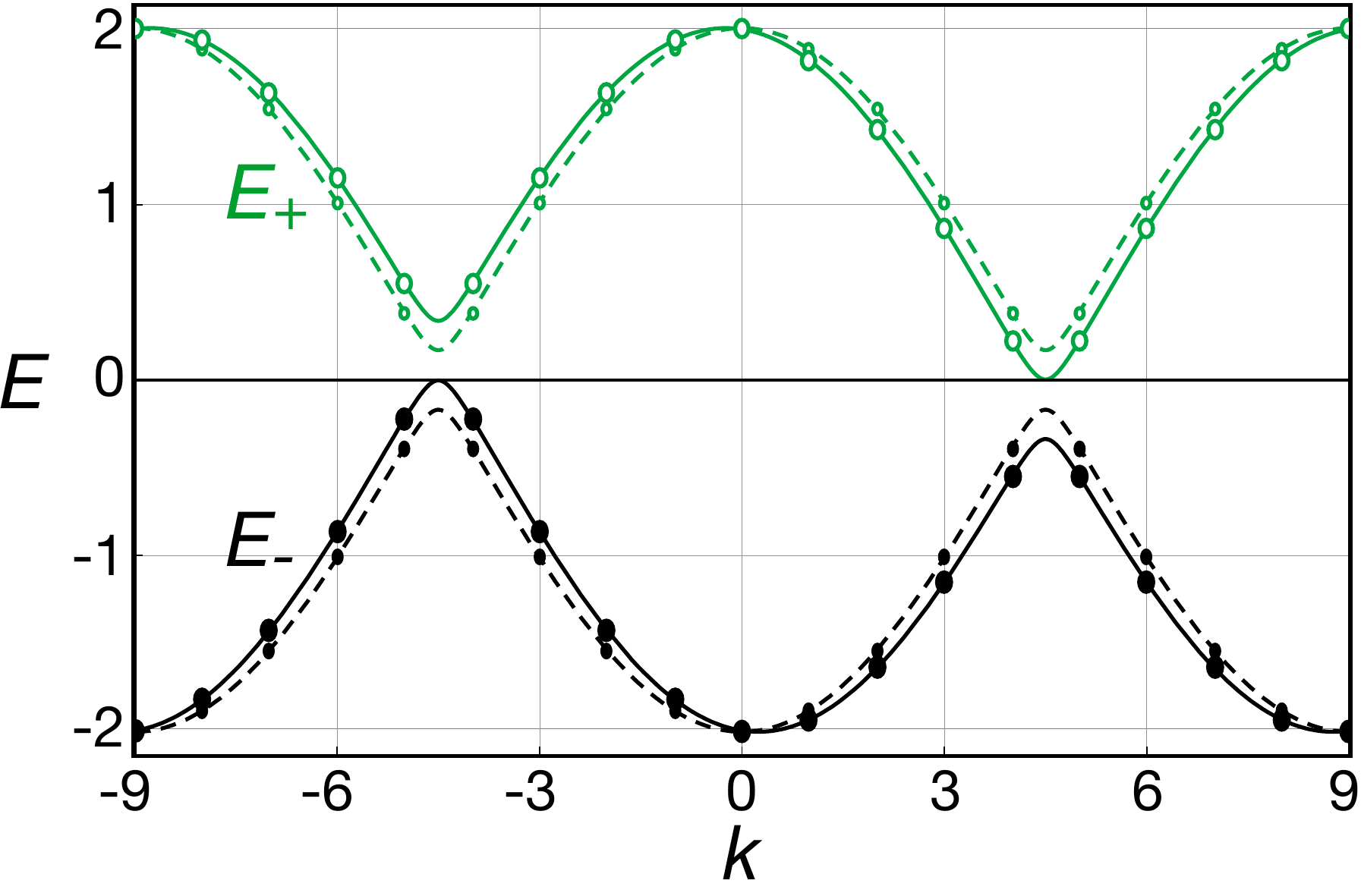}
\caption
{Energy dispersion of a ring with an order parameter $\Delta=0.22t$ for 
$\phi=0$ (dashed line) and $\phi=\phi_c\approx 0.24t$ (solid line), where the 
indirect energy gap closes. The filled (empty) circles represent occupied 
(unoccupied) $k$-states for a ring with $N=18$. The asymmetry for $\pm k$ scales with $1/R$.}
\label{Fig0}
\end{figure}

\begin{figure}[t]
\centering
\includegraphics[width=84mm]{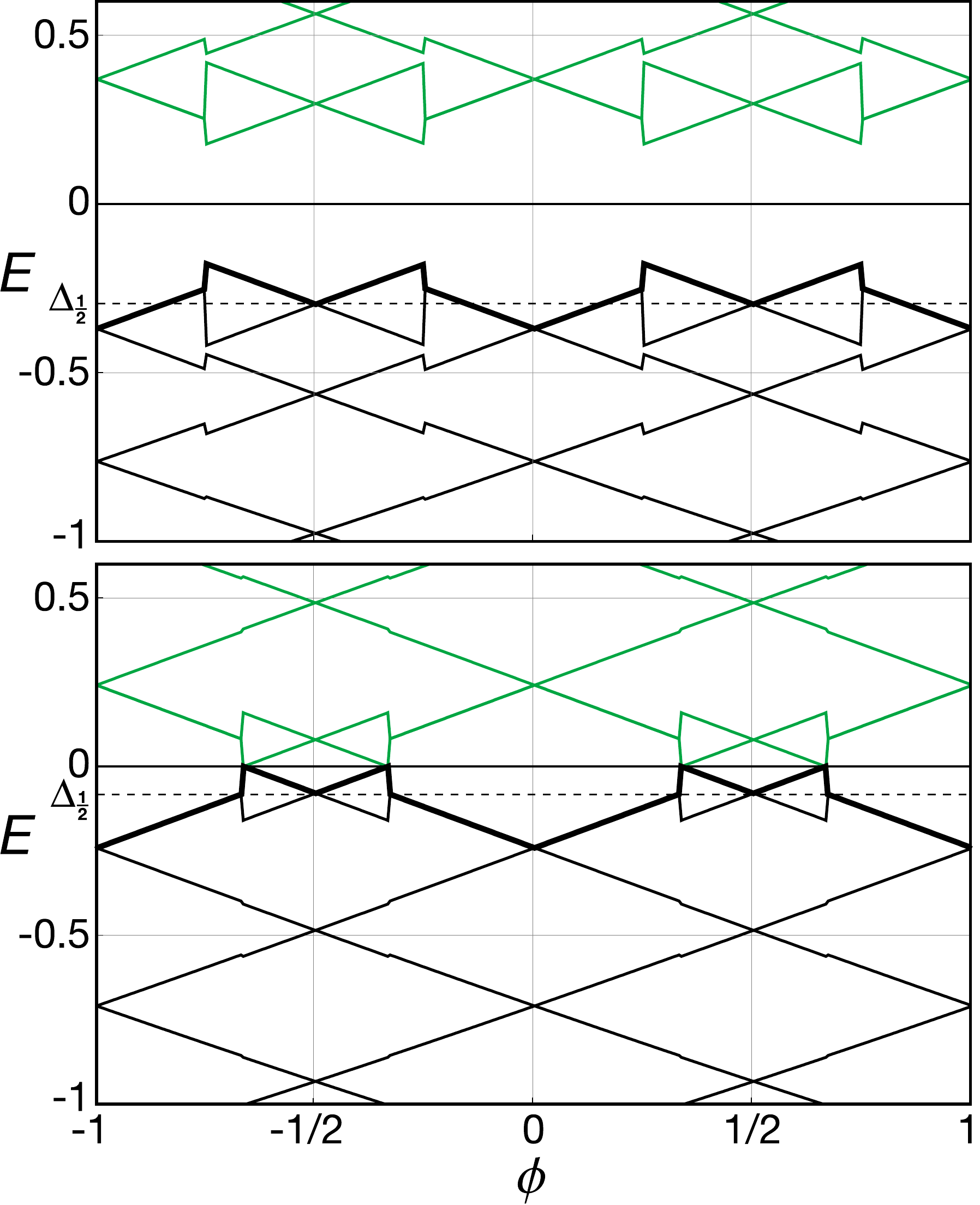}
\caption{\small
{Eigenenergies $E_\pm(k,\phi)$ (\ref{s10}) as a function of flux $\phi$ for $N=26$ and a 
self-consistently calculated order parameter $\Delta$; black lines: $E_-(k,
\phi)$, green lines: $E_+(k,\phi)$. Upper panel: \textquotedblleft large 
gap\textquotedblright regime ($V=1.9t$, $\Delta_{1/2}\approx0.30t$); Lower panel: 
\textquotedblleft small gap\textquotedblright regime ($V=1.1t$, $\Delta_{1/2}
\approx 0.08t$). Superconductivity occurs only in the odd-$q$  
sectors for $V=1.1t$ (see Fig.~\ref{Fig40}). The bold line marks the highest 
occupied state for all $\phi$. For the definition of $\Delta_{1/2}$ see text after Eq.~(\ref{s10.1}).} }
\label{Fig00}
\end{figure}

We assume that the unique ground state of the superconducting condensate is 
characterized by a single integer quantum number $q(\phi)$. 
For rings larger than the coherence length, 
the $q$-number of 
the ground state advances to the next integer whenever $\phi$ crosses the flux
values $(2n-1)/4$, $n\in\mathbb Z$, i.e.,  $q(\phi)=\mbox{floor}\left(2\phi
+1/2\right)$, where $\mbox{floor}(x)$ is the largest integer smaller than $x$. Discreteness of the energy levels shifts 
the increment in $q$ slightly according to the energy 
difference of even-$q$ and odd-$q$ states. 
Disregarding this small shift in a first approach (see comment at the end of  this section), we take $\Delta_k(q)$ of the form
\begin{equation}
\Delta_k(x)=\delta(x-q(\phi))\Delta_{k}.
\label{s7}
\end{equation}
For $s$-wave pairing, which is the only Cooper-pair state possible in a 
strictly $1D$ system, $\Delta_{k}\equiv\Delta$ is constant. With this ansatz, 
we determine the Green's function from Eq.~(\ref{s4}) as
\begin{equation}
G(k,\omega_n)=\frac{-i\hbar\omega_n-\epsilon_{-k-\phi+q}}{(i\hbar\omega_n-E_+(k,\phi))(i\hbar\omega_n-E_-(k,\phi))},
\label{s9}
\end{equation}
where the two energy branches $E_\pm(k,q)$ are given by
\begin{equation}
E_{\pm}(k,\phi)=\frac{\epsilon_{k-\phi}-\epsilon_{-k-\phi+q}}{2}\pm \sqrt{
\Delta^2+\epsilon^2(k,\phi)}
\label{s10}
\end{equation}
with $\epsilon(k,\phi)=(\epsilon_{k-\phi}+\epsilon_{-k+q-\phi})/2$. The energies $E_\pm(k,\phi)$ are plotted in Fig.~\ref{Fig0} as a function of $k$. The upper 
($E_+$) and the
lower branch ($E_-$) are separated by an indirect energy gap, 
which closes at a critical value $\Delta_c$. For finite flux the dispersion is
asymmetric with respect to an inversion in the angular momentum $k\rightarrow -k$
(see Fig. \ref{Fig0}), and this asymmetry induces a finite supercurrent. In 
the \textquotedblleft small gap" regime $\Delta<\Delta_c$, both $E_+(k,\phi)$ 
and $E_-(k,\phi)$ can be positive or negative, whereas in the 
\textquotedblleft large gap" regime $\Delta>\Delta_c$, $E_{+}(k,\phi)>0$ and 
$E_-(k,\phi)<0$ for all $k$ and $\phi$ (see Fig. \ref{Fig00}(a)).
Close to $E_F$, $E_\pm(k,\phi)$ simplifies to
\begin{equation}
E_\pm(\pm k,\phi)\approx \mp\frac{t}{R}(2\phi-q)\pm\sqrt{\Delta^2+l(t/R)^2}
\label{s10.1}
\end{equation}
where $k>0$ and $l=1$ for even $q$ and $l=0$ for odd $q$. The maximum direct energy gap in the even-$q$ sectors is therefore $\Delta_0 =\sqrt{\Delta^2+(t/R)^2}$, whereas in the odd-$q$ sectors it is $\Delta_{\frac{1}{2}}=\Delta$. Eq.~(\ref{s10.1}) shows that the shift of the eigenenergies scales with the ring size as $1/R$ in the \textquotedblleft small gap" regime. 

By inserting $G(k,\omega_n)$ into the Gor'kov equation (\ref{s6}), one finds  for 
the anomalous Green's function
\begin{equation}
F(k,k-q,\omega_n)=\frac{\Delta(\phi)}{(i\hbar\omega_n-E_+(k,\phi))(i\hbar\omega_n-E_-(k,\phi))}.
\label{s9.9}
\end{equation}
For a momentum independent pairing interaction $V_{kk'}\equiv V$ we
obtain the self-consistency equation for $\Delta(\phi)$ from Eq.(\ref{s011.1}) by 
summation over $\omega_n$
\begin{equation}
\frac{1}{N}\sum_k\frac{f(E_-(k,\phi))-f(E_+(k,\phi))}{2\sqrt{\Delta(\phi)^2+
\epsilon^2(k,\phi)}}=\frac{1}{V},
\label{s9.10}
\end{equation}
where $f(E)$ denotes the Fermi distribution function.
Instead of lowering the temperature we explore below the transition into the 
superconducting state at zero temperature by increasing the pairing 
interaction strength $V$. 

\begin{figure}[tb]
\centering
\includegraphics[width=84mm]{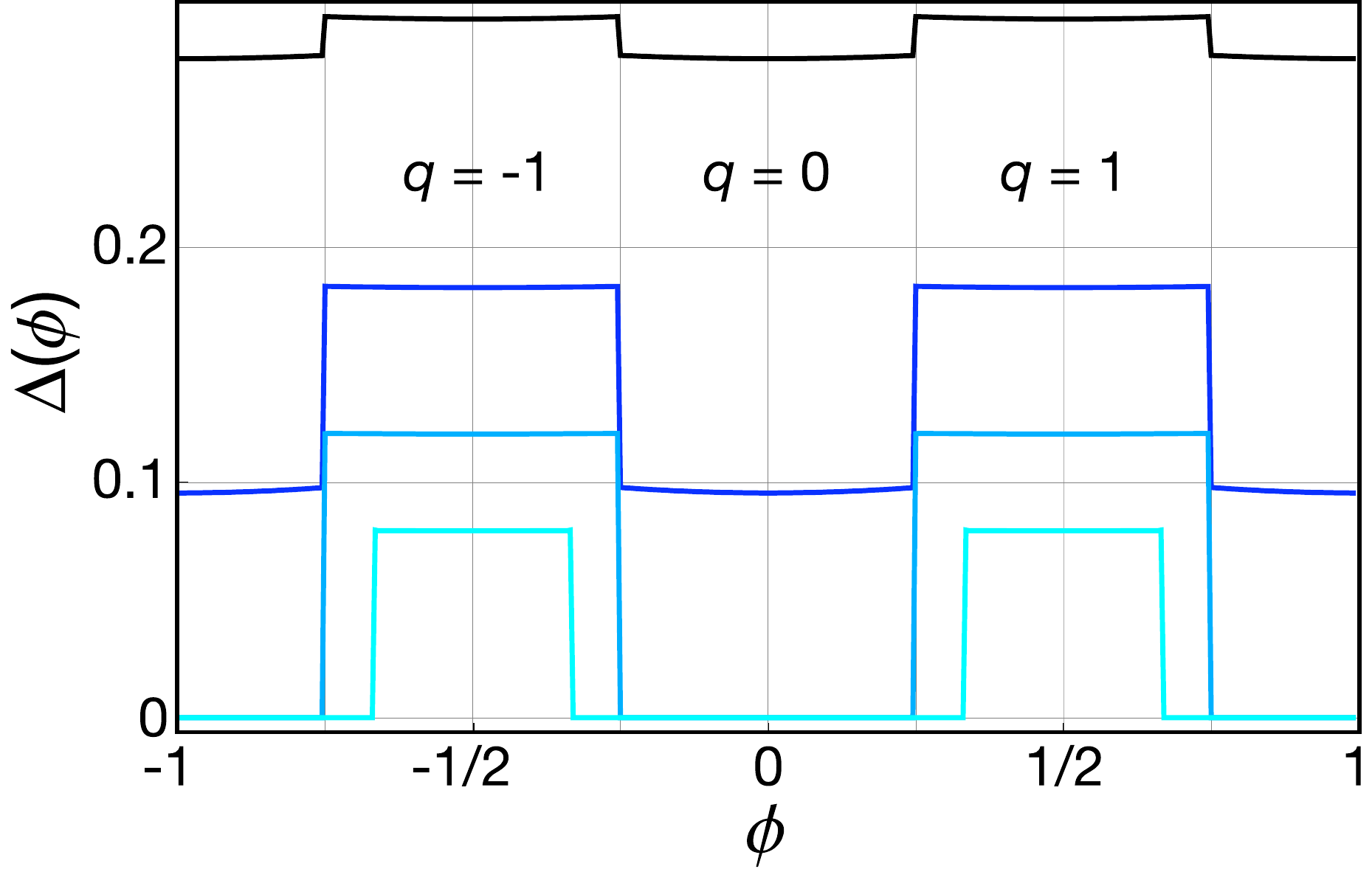}
\caption{
{Solution of the self-consistency equation (\ref{s9.10}) for different values 
of the pairing energy $V$ at $T=0$. From top to bottom: $V=1.9t,\, 1.6t,\, 1.35t,\, 
1.1t$. 
}}
\label{Fig40}
\end{figure}

\begin{figure*}[htb]
\centering
\subfigure[$\ V=0.0t$]{\includegraphics[height=40mm]{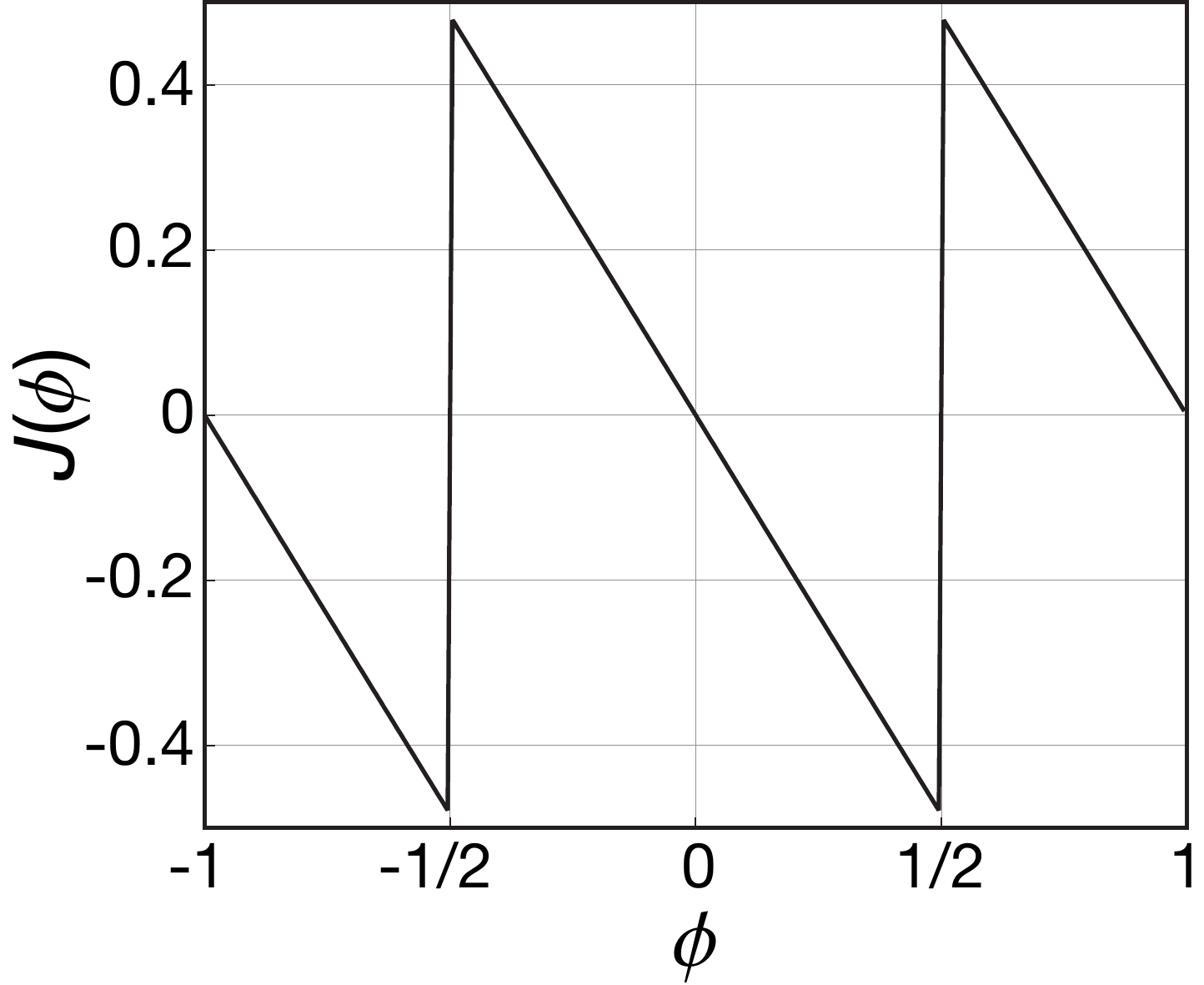}}
\subfigure[$\ V=0.5t$]{\includegraphics[height=40mm]{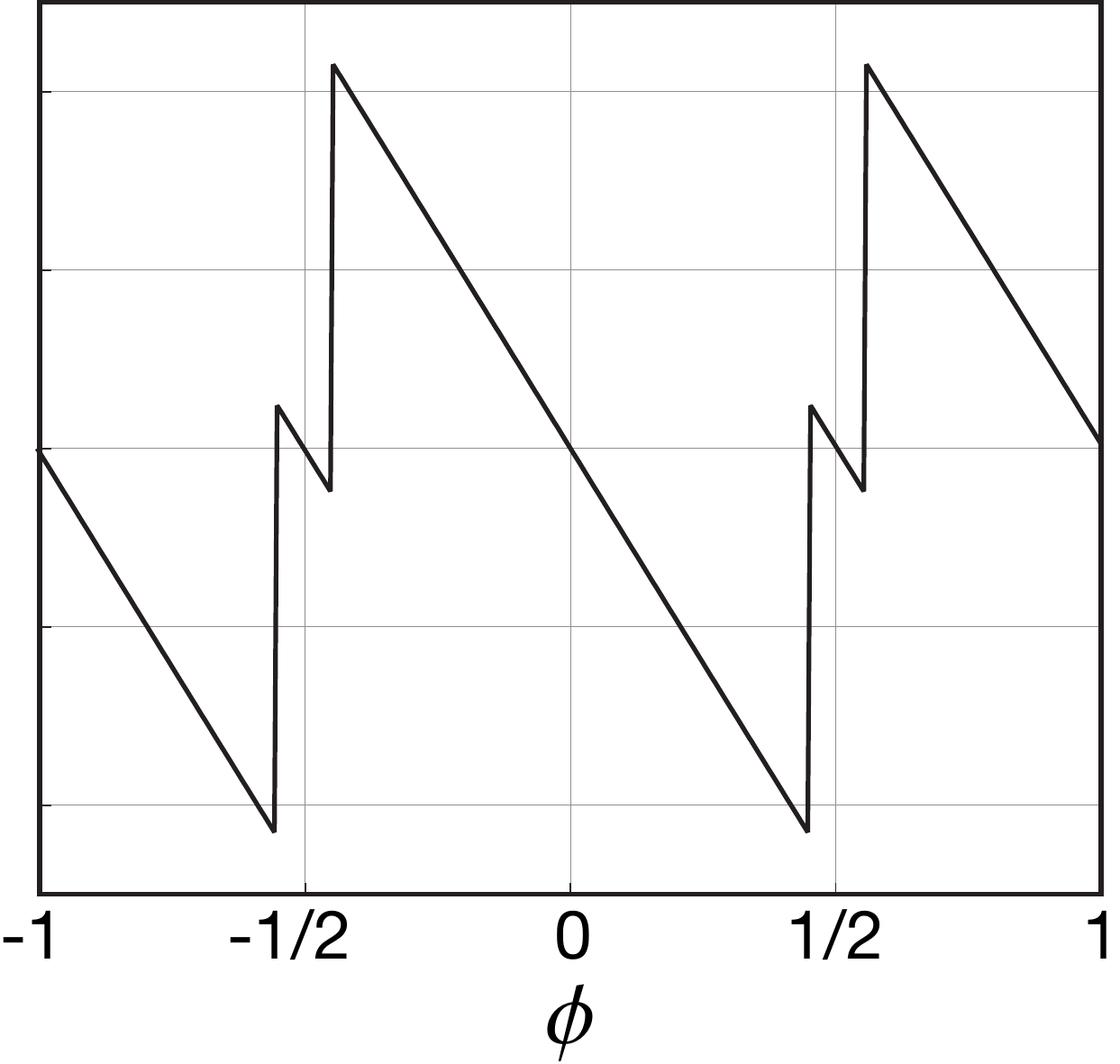}}
\subfigure[$\ V=1.1t$]{\includegraphics[height=40mm]{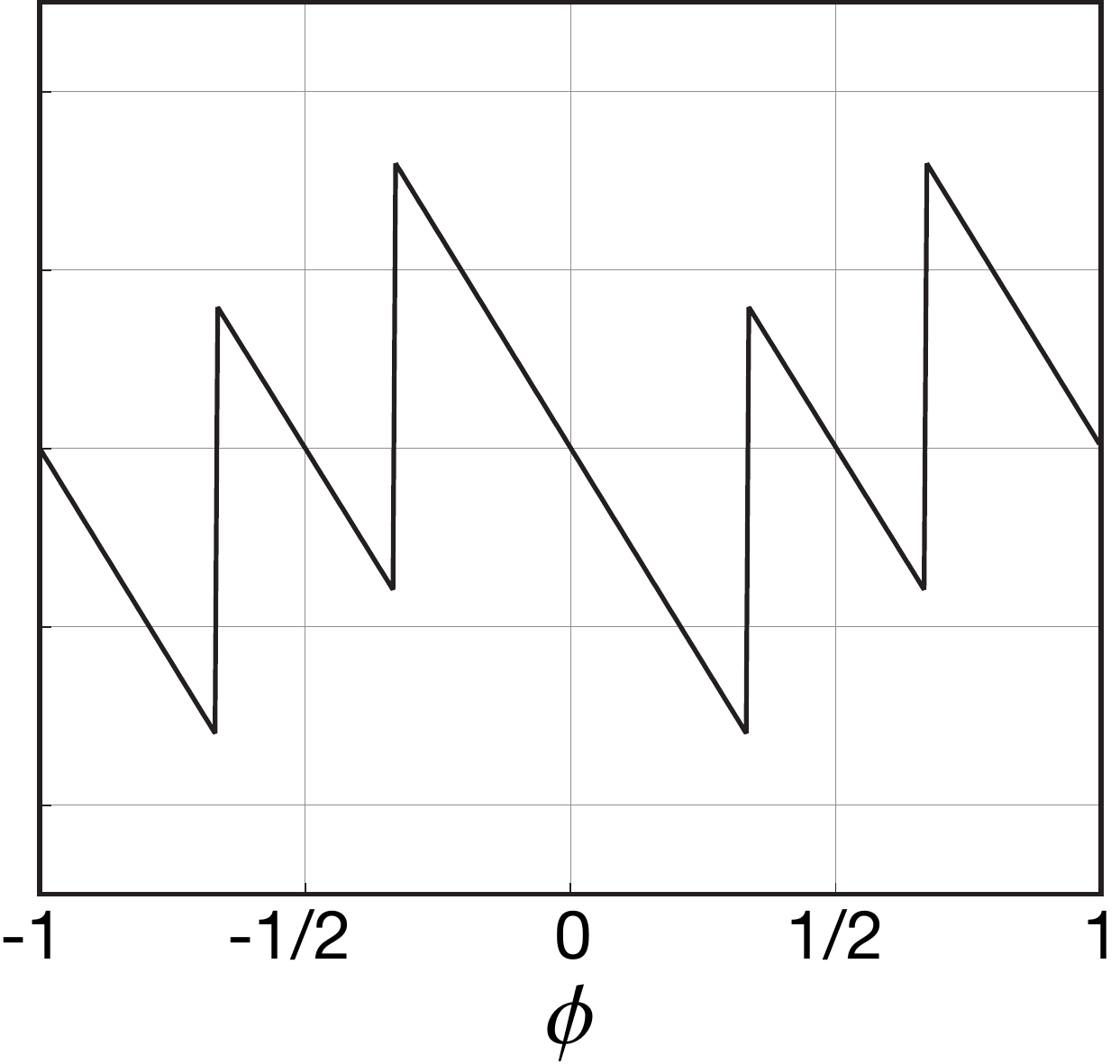}}
\subfigure[$\ V=1.9t$]{\includegraphics[height=40mm]{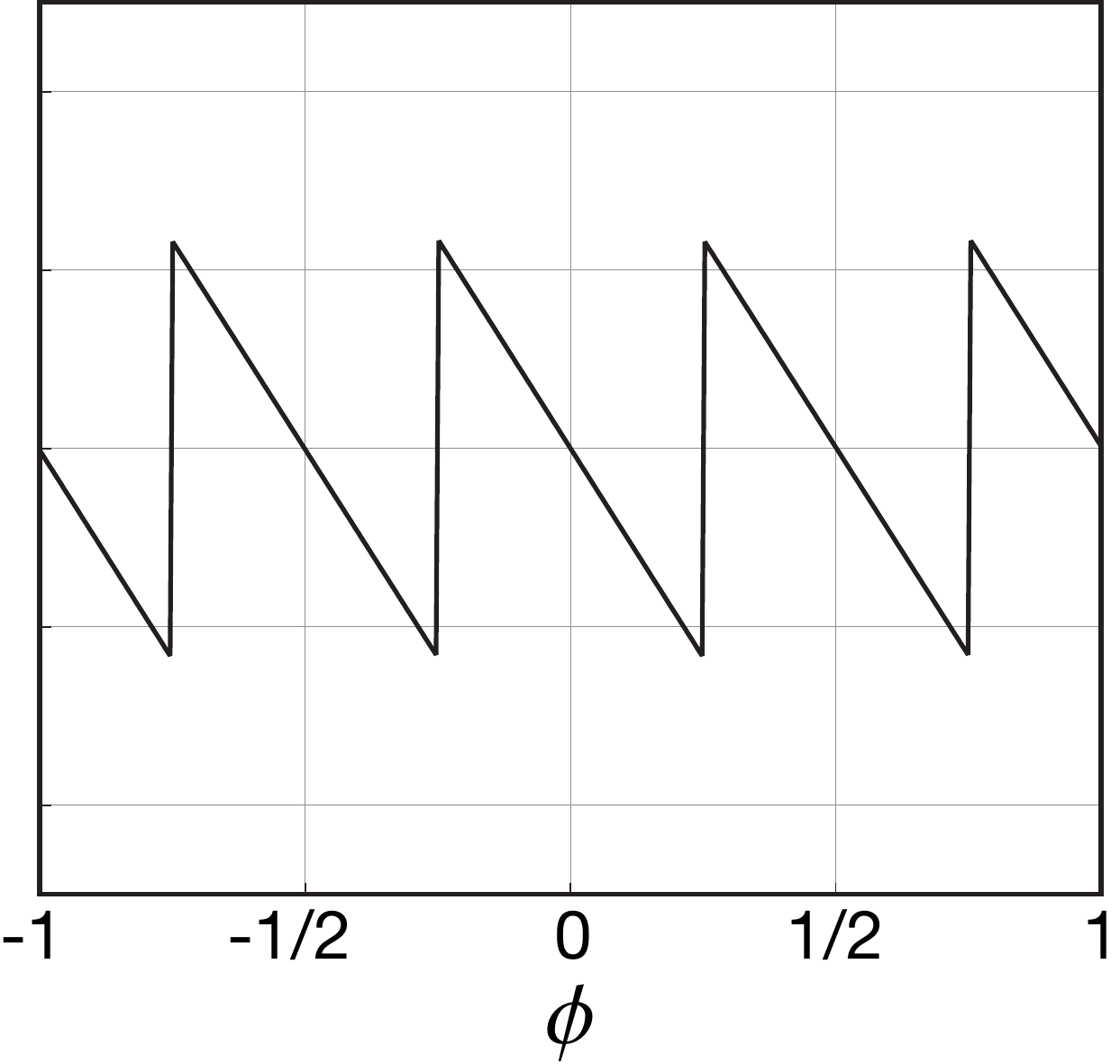}}
\caption{
Crossover from the $hc/e$-periodic normal persistent current to the $hc/2e$-periodic 
supercurrent in a ring with $N=26$ at $T=0$. For this ring size $\Delta_c
\approx0.24t$. The discontinuities occur where the $\phi$-derivative of the 
highest occupied state energy changes sign.}
\label{Fig3}
\end{figure*}

The flux $\phi$ affects the solution of the gap equation (\ref{s9.10}) for 
$\Delta$ in two ways. For small-size rings the magnitude of $\Delta$ is mainly 
controlled by the energy of the level closest to $E_F$. If the quantity 
$\delta_\phi=\min_k|\epsilon(k,\phi)-E_F|>0$, a solution of Eq.~(\ref{s9.10}) 
exists only above a threshold value of the pairing interaction. In the 
even-$q$ sectors, this is the case for all values of $\phi$, whereas in all 
odd-$q$ sectors a flux value $\phi$ exists, for which $\delta_\phi=0$ and 
Eq.~(\ref{s9.10}) has a solution for all $V>0$ (cf. Fig.~\ref{Fig40}). This is 
a consequence of the discreteness of the energy levels. In the strong coupling 
regime $V\gg t$, $\Delta$ is modulated only slightly by the flux. For weak 
coupling $V\approx t$, a solution 
$\Delta_{\frac{1}{2}}<\Delta_c$ is possible in the 
\textquotedblleft small gap" regime, where $\Delta_{\frac{1}{2}}$ denotes the order 
parameter at half-integer flux values. In this case the energy gap closes at a 
critical flux $\phi_c$ in the odd-$q$ sectors and $E_+(k,\phi)$ turns negative 
for the level closest to $E_F$ (see Fig.~\ref{Fig00}). Thus the dominant term 
in the sum of Eq.~(\ref{s9.10}) switches sign and the solution for $\Delta$ 
vanishes discontinuously. This is equivalent to a breaking of the Cooper pair 
closest to $E_F$, which provides the main contribution to the condensation energy \cite{bagwell:94,bardeen:62}.
These features for the solution of the self-consistency equation are special 
for strictly 1$D$ rings. In these rings superconductivity is destroyed for 
velocities of circulating Cooper pairs exceeding the Landau critical velocity, 
which is approached at $\phi=\phi_c$ \cite{zagoskin}.

With the discrete lattice gradient $\nabla_if(i)=\frac{1}{2}\left[f(i+1)-f(i-1)
\right]$, the current is obtained from
\begin{align}
J(\phi)=\left.\frac{-te}{\hbar}(\nabla_i\!-\!\nabla_j)G(i\!-\!j)e^{i\varphi_{ij}}
\right|_{i=j}\!=\frac{e}{h}\sum_{k}\frac{\partial\epsilon_{k}}{\partial k}n(k),
\label{s13}
\end{align}
where $n(k)=k_BT\sum_n G(k,\omega_n)$ is the momentum distribution function. The result is shown in Fig.~\ref{Fig3}. For $V=\Delta=0$ one recovers the 
$hc/e$-periodic saw-tooth pattern for the normal persistent current as 
discussed in \cite{cheung:88}. With increasing $\Delta$, new linear sections 
appear continuously. These are the sections where the order parameter is 
finite in the \textquotedblleft small-gap" regime \cite{czajka:05}. The 
occupied state closest to $E_F$ contributes dominantly to the current, because 
all other contributions tend to almost cancel in pairs. The discontinuities of the 
current occur where the $\phi$ derivative of the energy of the highest 
occupied state switches sign (see~Fig.~\ref{Fig00}). These linear sections 
increase with increasing $\Delta$; once they extend to a
range $hc/2e$ upon 
reaching the \textquotedblleft large gap" regime, the current becomes strictly 
$hc/2e$-periodic.

We obtain further insight into the mechanisms, which determine the current 
periodicity, by analyzing $\Delta_c$. According to Eq.(\ref{s10.1}), close to $E_F$, the maximum 
energy shift is $t/(2R)$, and the condition for a direct energy gap (or 
$E_+(k,\phi)>0$ for all $k$, $\phi$) and an $hc/2e$-periodic current pattern is 
therefore $\Delta>\Delta_c=t/(2R)$. The corresponding critical ring radius 
is $R_c=t/(2\Delta)$.

It is instructive to compare $R_c$ with the BCS coherence length 
$\xi_0=\hbar v_F/(\pi\Delta)$, where $v_F$ is the Fermi velocity and $\Delta$ 
the BCS order parameter at $T=0$. On the lattice we identify $v_F=\hbar k_F/m$ 
with $k_F=\pm\pi/2a$ and $m=\hbar^2/(2a^2t)$; $a$ is the lattice constant. 
Setting the length unit $a=1$ we obtain $\xi_0=t/\Delta$ and thus $2R_c=\xi_0$. This 
signifies that the current response of a superconducting ring smaller than 
the coherence length is generally $hc/e$-periodic \cite{loder:07}. In these rings the Cooper 
pair wave function is delocalized around the ring.

A second fundamental effect, which manifestly breaks the $hc/2e$-periodicity, is the offset
of the transition from even to odd center of mass angular momenta $q$ with respect to  evenly 
spaced flux values $(2n-1)\, hc/4e$. This small offset was already observed in our previous 
numerical evaluations for  $d$-wave loops \cite{loder:07}. Vakaryuk \cite{vakaryuk:08} has traced this shift to the
 dependence of the internal energy of Cooper pairs on the center of mass state. For a 
 BCS-model superconductor, this effect is fully incorporated in the Bogoliubov - de Gennes (BdG)
  evaluation of Ref.~\onlinecite{loder:07}
 although the quasiparticle-like presentation introduces a different perspective. In the discussion in this section we
 disregarded the offset for the 1$D$ rings in order to focus on the aspects related to the opening of
 an indirect gap. In Sec.~III we include the offset consistently in the BdG evaluation of the multi-channel annulus.

It is worthwhile to note that the condition $\Delta>\Delta_c$ (or $R>R_c$) only refers to the periodicity of 
the supercurrent. It does not guarantee an $hc/2e$-periodicity of the order 
parameter $\Delta$ or the total energy, but only of their derivatives. These 
quantities need a continuous energy spectrum with degeneracies for flux values 
which are multiples of $hc/2e$ \cite{schrieffer, brenig:61}. 

\section{Multichannel ring: annulus}\label{sec:annulus}

\begin{figure}[b]
\centering
\includegraphics[width=54mm]{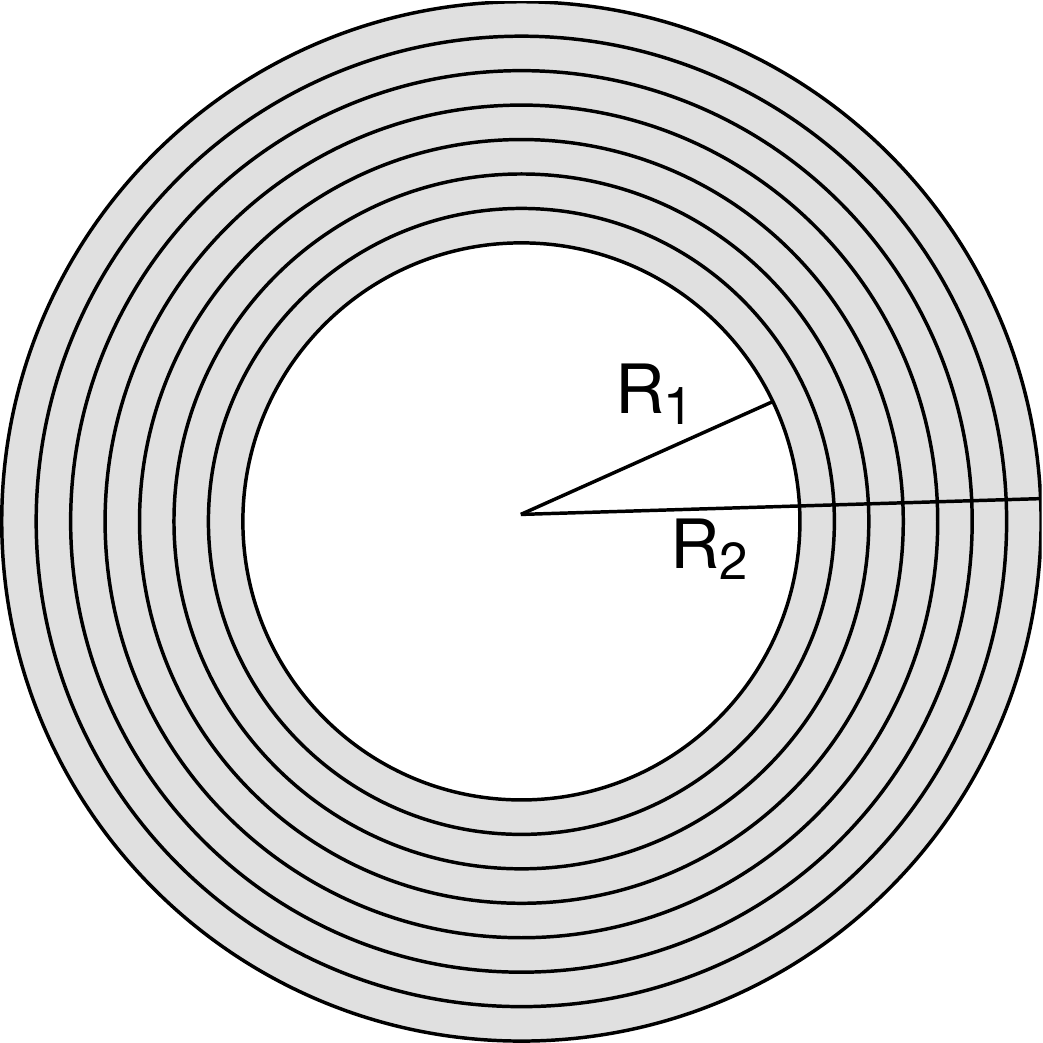}
\caption
{Annulus with inner radius $R_1$ and outer radius $R_2$. For a magnetic flux threading the interior of the annulus, the radial part of the Bogoliubov - de Gennes equations is solved numerically with a discretized radial coordinate.}
\label{Fig7}
\end{figure}

In this section we describe a superconducting loop of finite width as shown in Fig.~\ref{Fig7} with an inner radius $R_1$ and an outer radius $R_2$. For such an annulus, we choose a continuum approach on the basis of the Bogoliubov - de Gennes (BdG) equations. For integer and half-integer flux values, these equations can be solved analytically, as we show in subsection~A. For an arbitrary magnetic flux, we discuss a numerical solution in subsection~B.

Consider the BdG equations for spin singlet pairing
\begin{equation}
\begin{split}
E_{\bf n}u_{\bf n}({\bf r})&=\!\left[\frac{1}{2m}\!\left(i\hbar\bm\nabla+\frac{e}{c}{\bf A({\bf r})}\right)^2\!-\mu\right]\!u_{\bf n}({\bf r})\!+\Delta\,v_{\bf n}({\bf r})\\
E_{\bf n}v_{\bf n}({\bf r})&=-\!\left[\frac{1}{2m}\!\left(i\hbar\bm\nabla-\frac{e}{c}{\bf A({\bf r})}\right)^2\!-\mu\right]\!v_{\bf n}({\bf r})\!+\Delta^{\!*}u_{\bf n}({\bf r})
\end{split}\!,
\label{1}
\end{equation}
with the self-consistency condition (gap equation) for the order parameter $\Delta({\bf r})$:
\begin{align}
\begin{split}
\Delta({\bf r})=V\sum_{\bf n}u_{\bf n}({\bf r})v^*_{\bf n}({\bf r})\tanh\left(\frac{E_{\bf n}}{2T}\right),
\end{split}
\label{34}
\end{align}
where $V$ is the local pairing potential. For an annulus of finite width we separate the angular part of the quasi-particle wave functions $u_{\bf n}({\bf r})$, $v_{\bf n}({\bf r})$ using polar coordinates ${\bf r}=(r,\varphi)$ and the ansatz
\begin{equation}
\begin{split}
u_{\bf n}(r,\varphi)&=u_{\bf n}(r)e^{\frac{i}{2}(k+q)\varphi}\\
v_{\bf n}(r,\varphi)&=v_{\bf n}(r)e^{\frac{i}{2}(k-q)\varphi}
\end{split},
\label{2}
\end{equation}
where $k$ and $q$ are either both even or both odd integers. Thus  $\hbar k$ is the angular momentum as for the $1D$ ring and ${\bf n}=(k,\rho)$ with the radial quantum number $\rho$.
The order parameter factorizes into $\Delta(r,\varphi)=\Delta(r)e^{iq\varphi}$ where the radial component
\begin{align}
\Delta(r)=V\sum_{\bf n}u_{\bf n}(r)v^*_{\bf n}(r)\tanh\left(\frac{E_{\bf n}}{2T}\right)
\label{38}
\end{align}
is real. For a magnetic flux $\Phi$ threading the interior of the annulus we choose the vector potential ${\bf A}(r,\varphi)={\bf e}_\varphi\,\Phi/(2\pi r)$, where ${\bf e}_\varphi$ is the azimuthal unit vector. With $\phi=(e/hc)\Phi$ and
\begin{align}
\left(-i\bm\nabla\pm\frac{\phi}{r}{\bf e}_\varphi\right)^2=-\frac{1}{r}\partial_r(r\partial_r)+\frac{1}{r^2}(-i\partial_\varphi\pm\phi)^2
\label{3}
\end{align}
the BdG equations therefore reduce to radial differential equations for $u_{\bf n}(r)$ and $v_{\bf n}(r)$:
\begin{equation}
\begin{split}
E_{\bf n}\,u_{\bf n}(r)&=\!-\!\!\left[\frac{\hbar^2}{2m}\frac{\partial_r}{r}(r\partial_r)\!-\!\frac{\hbar^2l_u^2}{2mr^2}\!+\!\mu\right]\!u_{\bf n}(r)\!+\Delta(r)v_{\bf n}(r)\\
E_{\bf n}\,v_{\bf n}(r)&=\!\left[\frac{\hbar^2}{2m}\frac{\partial_r}{r}(r\partial_r)\!-\!\frac{\hbar^2l_v^2}{2mr^2}\!+\!\mu\right]\!v_{\bf n}(r)\!+\Delta(r)u_{\bf n}(r)
\end{split}\!,
\label{4}
\end{equation}
with the canonical angular momenta
\begin{align}
\hbar l_u&=\frac{\hbar}{2}(k+q-2\phi),\\
\hbar l_v&=\frac{\hbar}{2}(k-q+2\phi).
\label{5}
\end{align}
The number $q$ plays the same role as in the previous section. Here we choose $q$ for each value of the flux to minimize the total energy of the system. The flux for which $q$ changes to the next integer can therefore deviate from the values $(2n-1)/4$, where we fixed the change of $q$ for the $1D$ model.
\subsection{Hankel-Function Ansatz}
A natural choice of an ansatz for the solutions of the coupled differential equations (\ref{4})  are linear combinations of the Hankel functions $H^{(1)}_l$ and $H^{(2)}_l$, since they are individually solutions of the uncoupled  equations (\ref{4}) for $\Delta(r)=0$:
\begin{align}
\left(\frac{1}{r}\partial_r(r\partial_r)-\frac{l^2}{r^2}\right)H^{(1,2)}_l(\gamma r)=\gamma^2H^{(1,2)}_l(\gamma r).
\label{6}
\end{align}
We therefore take $u_{\bf n}(r)$ and $v_{\bf n}(r)$ of the form:
\begin{align}
u_{\bf n}(r)=u_{\bf n}\left[H^{(1)}_{l_u}(\gamma^u_{\bf n}r)+c^u_{\bf n}H^{(2)}_{l_u}(\gamma^u_{\bf n}r)\right],
\label{7.5}\\
v_{\bf n}(r)=v_{\bf n}\left[H^{(1)}_{l_v}(\gamma^v_{\bf n}r)+c^v_{\bf n}H^{(2)}_{l_v}(\gamma^v_{\bf n}r)\right].
\label{7}
\end{align}
The equations (\ref{4}) then become
\begin{equation}
\begin{split}
E_{\bf n}\,u_{\bf n}(r)]&=-\left[\frac{\hbar^2}{2m}\left(\gamma^u_{\bf n}\right)^2+\mu\right]u_{\bf n}(r)+\Delta(r)v_{\bf n}(r)\\
E_{\bf n}\,v_{\bf n}(r)&=\left[\frac{\hbar^2}{2m}\left(\gamma^v_{\bf n}\right)^2+\mu\right]v_{\bf n}(r)+\Delta(r)u_{\bf n}(r)
\end{split},
\label{8}
\end{equation}
The coefficients $\gamma^\alpha_{\bf n}$ and $c^\alpha_{\bf n}$ with $\alpha=u,v$ are fixed by the open boundary conditions, for which $u_{\bf n}(r)$ and $v_{\bf n}(r)$ vanish on the inner and outer boundaries of the annulus: $u_{\bf n}(R_1)=u_{\bf n}(R_2)=0$ and $v_{\bf n}(R_1)=v_{\bf n}(R_2)=0$. This generates the defining equations for $\gamma^\alpha_{\bf n}$ and $c^\alpha_{\bf n}$
\begin{align}
c^\alpha_{\bf n}=-\frac{H^{(1)}_{l_\alpha}(\gamma^\alpha_{\bf n}R_1)}{H^{(2)}_{l_\alpha}(\gamma^\alpha_{\bf n}R_1)}=-\frac{H^{(1)}_{l_\alpha}(\gamma^\alpha_{\bf n}R_2)}{H^{(2)}_{l_\alpha}(\gamma^\alpha_{\bf n}R_2)}.
\label{9}
\end{align}
For all integer and half-integer values of flux, $q=2\phi$ in the ground state, thus $l_u=l_v=k/2$. Assuming a constant order parameter $\Delta(r)=\Delta$, the $r$-dependence drops out from Eqs.~(\ref{8}) and we find the eigenvalues and eigenvectors of the usual BCS type
\begin{align}
E_{\bf n}=\sqrt{\left(\frac{\hbar^2}{2m}\gamma_{\bf n}^2-\mu\right)^2+\Delta^2}
\label{10}
\end{align}
with $\gamma_{\bf n}=\gamma_{\bf n}^u=\gamma_{\bf n}^v$ and
\begin{align}
u_{\bf n}&=\frac{1}{2}\left[1+\left(\frac{\hbar^2}{2m}\gamma_{\bf n}^2+\mu\right)/E_{\bf n}\right],\\
v_{\bf n}&=\frac{1}{2}\left[1-\left(\frac{\hbar^2}{2m}\gamma_{\bf n}^2+\mu\right)/E_{\bf n}\right].
\label{11}
\end{align}
These are the two distinct classes of superconducting states as discussed for the $1D$ loop: for integer flux values, $\Delta$ is given by summing over all even angular momenta $k$, whereas for half-integer flux values, $\Delta$ is obtained by summing over odd angular momenta.

For general values of magnetic flux, $l_u$ and $l_v$ are different and so are $\gamma^u_{\bf n}$ and $\gamma^v_{\bf n}$. The $r$-dependence of $u_{\bf n}(r)$ is therefore different from $v_{\bf n}(r)$ as contained in Eqs.~(\ref{7.5},\ref{7}). In App.~A we analyze the solution of the uncoupled Eqs.~(\ref{4}) for $\Delta=0$ and find that the eigenfunctions account for the flux induced Doppler shift by shifting their nodes closer together or further 
apart---most importantly, $u_{\bf n}(r)$ shifts its nodes in the opposite direction than does $v_{\bf n}(r)$. This implies that $u_{\bf n}(r)$ and $v_{\bf n}(r)$ with the ansatz of Eqs.~(\ref{7.5}) and (\ref{7}) cannot be solutions of the coupled Eqs.~(\ref{8}) for $\Delta(r)\neq0$.

Moreover, we show in App.~A for the limit of a thin annulus ($R_1\gg R_2-R_1$) that both the Doppler shift and the shift of the nodes of $u_{\bf n}(r)$ and $v_{\bf n}(r)$ are in leading order linear functions of $q-2\phi$. It is therefore not possible to find an approximate solution of Eqs.~(\ref{4}) that contains the effects of the Doppler shift but neglects the shift of the nodes. Consequently, we have to resort to a numerical solution of the radial component of the BdG equations.
\subsection{Self-Consistent Numerical Solution}
The numerical solution of Eqs.~(\ref{4}) is achieved by discretizing the interval $[R_1,R_2]$ for the radial coordinate $r$ into $M$ radii $r_i$, which defines the grid constant $a=(R_2-R_1)/M$. In this way we obtain for each angular momentum $\hbar k$ $M$ radial eigenstates (channels), which correspond to the $M$ eigenstates with the lowest eigenenergies $E_{\bf n}$ of the continuum model. On this set of $M$ radial coordinates, we use the symmetric discrete differential operators $\partial_if(r_i)=[f(r_{i+1})-f(r_{i-1})]/a$ and $\partial_i^2f(r_i)=[f(r_{i+1})+f(r_{i-1})-2f(r_i)]/a^2$. Inserting these discrete operators into Eqs.~(\ref{4}) and using $(1/r)\partial_rr\partial_r=(1/r)\partial_r+\partial_r^2$, one obtains the eigenvalue equation
\begin{align}
\begin{pmatrix}
\hat t+\hat\mu^u_k&\widehat\Delta\cr\widehat\Delta&-\hat t-\hat\mu^v_k
\end{pmatrix}
\begin{pmatrix}
{u}_{\bf n}\cr{v}_{\bf n}
\end{pmatrix}
=E_{\bf n}\begin{pmatrix}{u}_{\bf n}\cr{v}_{\bf n}\end{pmatrix}
\label{25}
\end{align}
where $u_{\bf n}$ and $v_{\bf n}$ are real and the operators $\hat t$, $\hat\mu_k^\alpha$, and $\widehat\Delta$ are defined through
\begin{multline}
\hat tu_{\bf n}(r_i)=t[u_{\bf n}(r_{i+1})+u_{\bf n}(r_{i-1})]\\
+t\frac{a}{r_i}[u_{\bf n}(r_{i+1})-u_{\bf n}(r_{i-1})],
\label{26}
\end{multline}
and
\begin{align}
\hat\mu^\alpha_ku_{\bf n}(r_i)&=t\left[\frac{a^2}{r_i^2}l_\alpha^2-2\right]u_{\bf n}(r_i),\\
\widehat\Delta u_{\bf n}(r_i)&=\Delta(r_i)u_{\bf n}(r_i),
\label{27}
\end{align}
where $t=\hbar^2/(2ma^2)$. A self-consistent solution of Eq.~(\ref{25}) and the gap equation
\begin{align}
\Delta(r_i)=V\sum_{\bf n}u_{\bf n}(r_i)v_{\bf n}(r_i)\tanh\left(\frac{E_{\bf n}}{2T}\right)
\label{35}
\end{align}
 is found iteratively.
The operator $\hat t$ consists of a symmetric and an antisymmetric part with respect to $r_{i-1}$ and $r_{i+1}$. In order to ensure that the eigenvalues of Eq.~(\ref{25}) are real, the prefactor of the second, antisymmetric term in Eq.~(\ref{26}) must be smaller or equal to the prefactor of the symmetric term, which means $M\geq(R_2-R_1)/2R_1$. This condition is fulfilled since $M= (R_2-R_1)/a > (R_2-R_1)/2R_1$.

Once the eigenfunctions of Eq.~(\ref{25}) are known, we obtain the current by evaluating the expectation value of the gauge invariant current operator \cite{bagwell:94}.
The expectation value $J(r)$ of the circulating current is found using a Bogoliubov transformation and the ansatz~(\ref{2}) in polar coordinates:
\begin{align}
J(r)=\frac{\hbar e}{m}\sum_{\bf n}\left[J^u_{\bf n}(r)f(E_{\bf n})-J^v_{\bf n}(r)f(-E_{\bf n})\right],
\label{33}
\end{align}
with
\begin{align}
\begin{split}
J_{\bf n}^\alpha(r)=&\frac{\hbar e}{m}\text{Im}\left[\alpha_{\bf n}^*(r,\varphi)\left(-\frac{i}{r}\partial_\varphi-\frac{\phi}{r}\right)\alpha_{\bf n}(r,\varphi)\right]\\
=&\frac{\hbar e}{m}\frac{l_\alpha}{r}\alpha_{\bf n}^2(r)
\end{split}
\label{32}
\end{align}
for $\alpha=u,v$. The contribution of each quasi-particle state to the total current is therefore determined by its angular velocity $l_\alpha$. The radial quantum number $\rho$ and the $\Delta$-dependence enter only through the occupation probability which is controlled by the eigenenergy $E_{\bf n}$. Further, the total energy of the system is given by
\begin{align}
\begin{split}
E=\frac{1}{M}\sum_{\bf n}E_{\bf n}\sum_i\left[u_{\bf n}^2(r_i)f(E_{\bf n})+v_{\bf n}^2(r_i)f(-E_{\bf n})\right].
\end{split}
\label{36}
\end{align}

\begin{figure}[t]
\centering
\includegraphics[width=84mm]{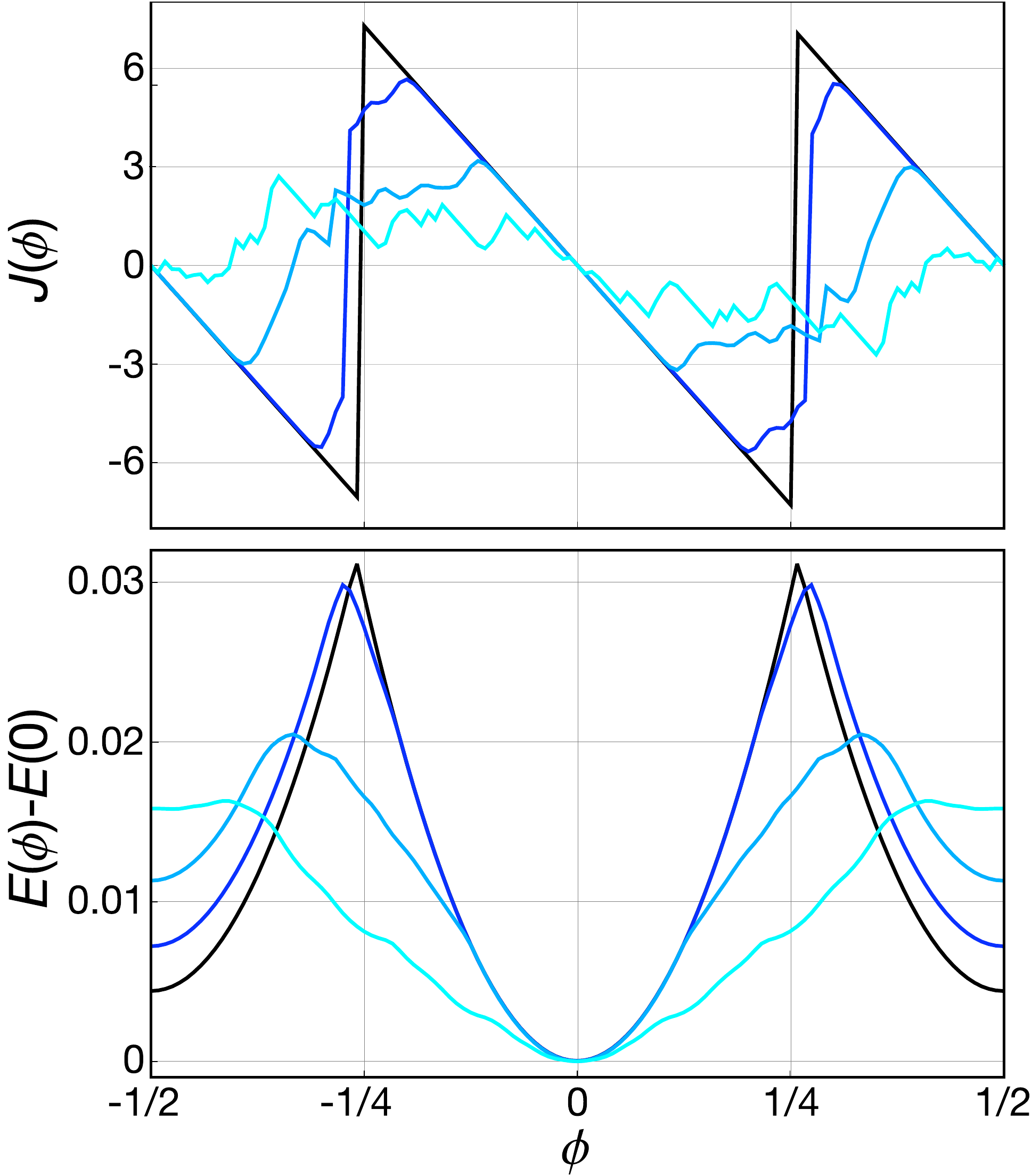}
\caption
{Non-self-consistent calculation of current and energy at $T=0$. The circulating current (upper panel) in an annulus with an inner radius $R_1=100a$ and an outer radius $R_2=150a$ is shown for fixed, $\phi$-independent $\Delta=0$ (light blue line), $\Delta=0.002t$ (blue line), $\Delta=0.004t$ (dark blue line), $\Delta=0.006t$ (black line).
The lower panel shows the difference between the total energy of the annulus as a function of $\phi$ and the total energy at zero flux for the same values for $\Delta$ as above.}
\label{Fig4}
\end{figure}

\begin{figure}[t]
\centering
\includegraphics[width=84mm]{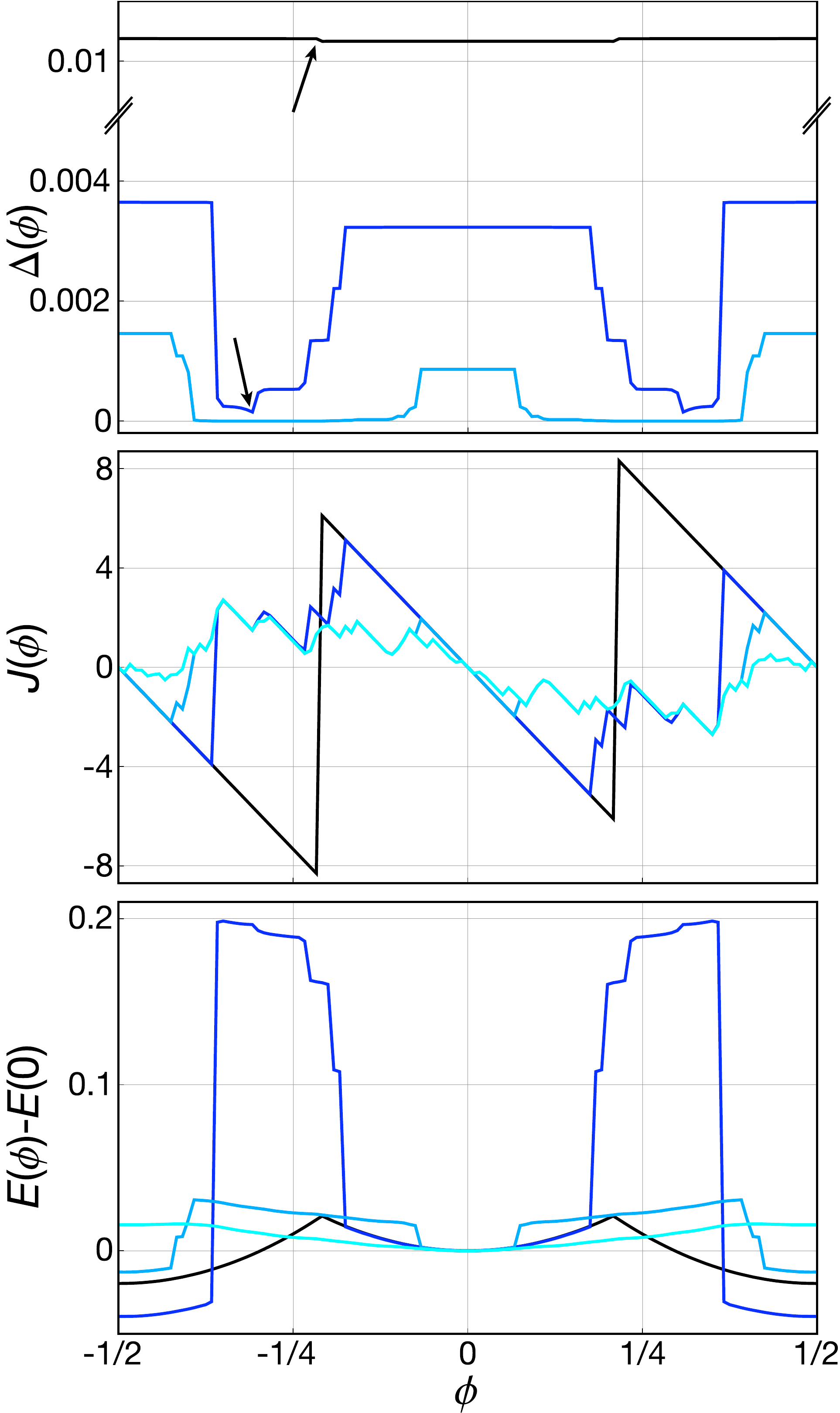}
\caption
{Self-consistent calculations for the same annulus as in Fig.~\ref{Fig4}. In addition, the top panel displays the self-consistent order parameter $\Delta$ as a function of $\phi$. The lines correspond to the pairing interaction $V=0$ (light blue line), $V=0.28t$ (blue line), $V=0.32t$ (dark blue line), $V=0.38t$ (black line). The black arrows mark the positions of the $q$-jump for $V=0.38t$ and $V=0.32t$.}
\label{Fig5}
\end{figure}

\subsection{Results}
The results of the non self-consistent calculations for the circulating current and the total energy at $T=0$ and for fixed values of $\Delta$ are displayed in Fig.~\ref{Fig4}. In the normal state ($\Delta=0$), there are $\sim\!M$ eigenstates close enough to $E_F$ to cross $E_F$ as a function of $\phi$, unlike in small $1D$ rings where only one state crosses $E_F$. For each crossing, a small jump appears in the current as a function of $\phi$. There is a larger jump at the value of $\phi$ where the energies of the even-$q$ and odd-$q$ states become degenerate and $q$ switches to the next integer. The shape of this function depends on the distribution of eigenenergies close to $E_F$ and therefore on microscopic details of the geometry of the annulus and the Fermi energy $E_F$. A finite $\Delta$ allows for a flux regime with direct energy gap and no crossings of $E_F$, thus in this regime the current is linear and the total energy quadratic in $\phi$. For the largest value ($\Delta=0.006t$) shown, there is a direct gap for all values of $\phi$. Even for this value of $\Delta$, the current and the energy are not exactly $hc/2e$-periodic because of the energy difference of the even and odd $q$ states in finite systems \cite{vakaryuk:08,loder:07}.

The introduction of self-consistency in $\Delta$ does not fundamentally change these basic observations (Fig.~\ref{Fig5}). The crossover is then controlled by the pairing interaction strength $V$, for which we chose such values as to reproduce the crossover from the normal state to a state with direct energy gap for all flux values. The order parameter $\Delta$ is now a function of $\phi$. If $\Delta(\phi=0)\lesssim 0.006t$ (cf.\ Fig.~\ref{Fig4}), the gap closes with $\phi$ and $\Delta$ decreases whenever a state crosses $E_F$. At these flux values we observe a sharp increase  in the total energy of the annulus. Unlike in $1D$, $\Delta$ does not drop to zero at the closing of the energy gap, but decreases stepwise.  In two or three dimensions, $\Delta$ remains finite beyond $\phi_c$ because it is stabilized by contributions to the condensation energy from pairs with relative momenta perpendicular to the direction of the current flow and the closing of the indirect energy gap does not destroy superconductivity \cite{bardeen:62, zagoskin}. Apart from these steps, the current (energy) shows the standard linear (quadratic) behavior. 

The offset of the $q$-jump is only relevant for values of $V$ for which $\Delta$ is finite for all $\phi$. In Fig.~\ref{Fig5}, the offset is clearly visible for the largest two values of $V$ (marked with black arrows). Its sign depends on the geometry of the annulus and the pairing interaction $V$--- the offset changes sign for increasing $V$ (cf.\ Ref.~\onlinecite{vakaryuk:08}). 

\begin{figure}[t]
\centering
\includegraphics[width=84mm]{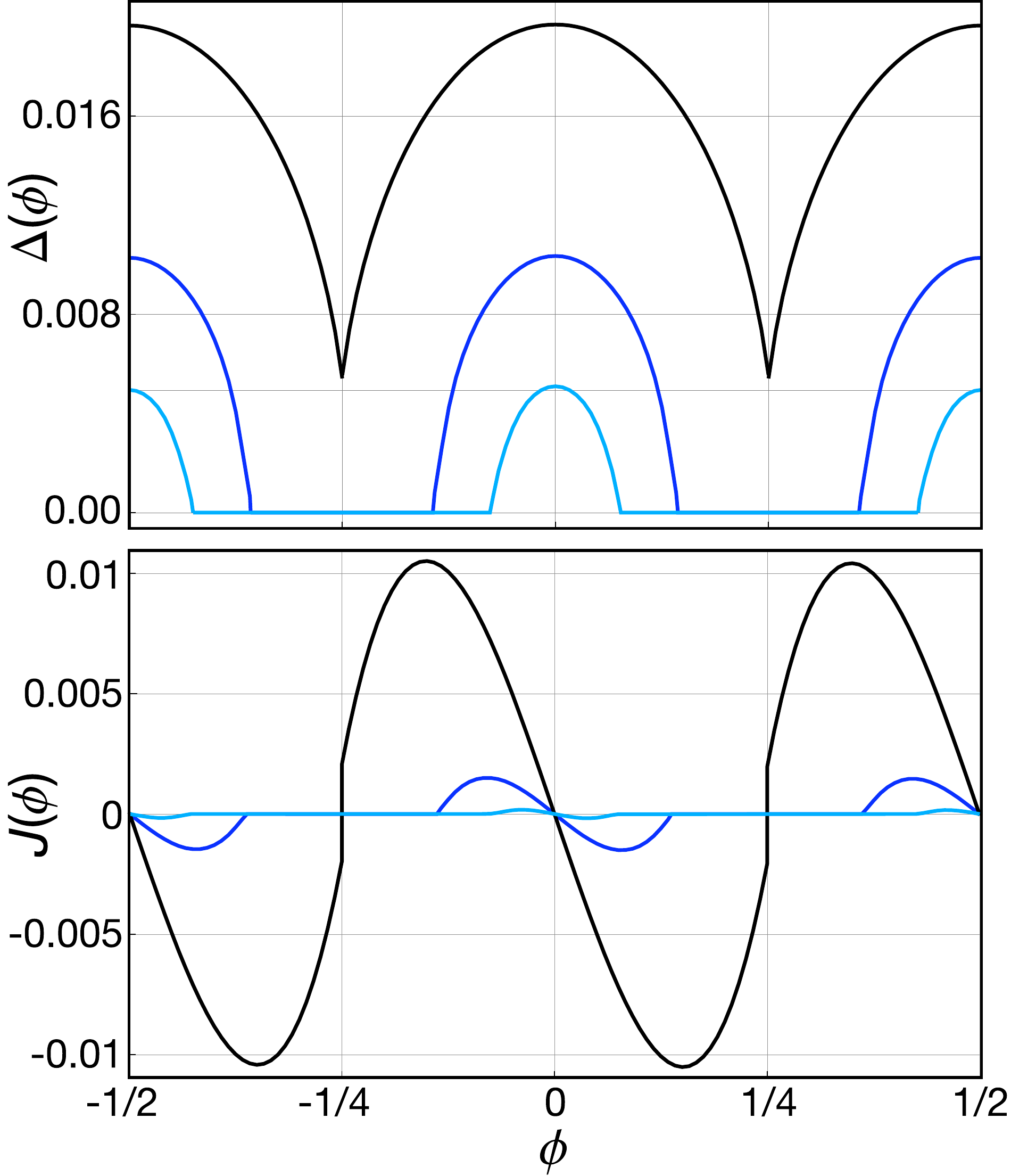}
\caption
{The order parameter $\Delta$ and the persistent current for the temperature driven transition from the normal to the superconducting state in an annulus with inner radius $R_1=30a$ and outer radius $R_2=36a$. The pairing intercation is $V=0.7\,t$, with a critical temperature of $T_c\approx0.0523\,t$  for zero flux. For these parameters $\Delta(T=0)\approx0.1t$. The lines correspond to the temperatures $T= 0.0513t$ (black line), $T= 0.0520t$ (dark blue line), $T= 0.0522t$ (blue line). Notice that $\Delta$ is slightly different for the flux values $\phi=0$ and $\phi=\pm1/2$.}
\label{Fig6}
\end{figure}

Experimentally more relevant is to control the  crossover through temperature. With the pairing interaction $V$ sufficiently strong to produce a $T=0$ energy gap much larger than the maximum Doppler shift, the crossover regime is reached for temperatures slightly below $T_c$. For the annulus described in Fig.~\ref{Fig6}, the crossover proceeds within approximately one percent of $T_c$. The crossover regime gets narrower for larger rings proportional to the decrease of the Doppler shift. In the limit of a quasi $1D$ ring of radius $R$ we can be more precise: If we define the crossover temperature $T^*$ by $\Delta(T^*)=\Delta_c$ and assuming $\Delta_c\ll\Delta$, we can use the Ginzburg-Landau form of the order parameter
\begin{align}
\frac{\Delta(T)}{\Delta(0)}\approx1.75\sqrt{1-\frac{T}{T_c}}
\label{41}
\end{align}
and obtain
\begin{align}
\frac{T_c-T^*}{T_c}\approx\frac{\Delta_c^2}{3.1\Delta(0)^2}=\frac{t^2}{12.4\Delta(0)^2R^2}=\frac{E_F^2}{3.1T_c^2R^2},
\label{37}
\end{align}
For a ring with a radius of 2500 lattice constants ($\approx10$ $\mu$m) and $\Delta(0)=0.01t$ ($\approx3$ meV) one finds the ratio $(T_c-T^*)/T_c\approx1.3\times10^{-4}$. This is in reasonable qualitative agreement with the experimental results of Little and Parks \cite{Little,Parks}, discussed also by Tinkham \cite{Tinkham}. Their theoretical prediction is similar to Eq.~(\ref{37}), up to a factor in which they include a finite mean free path. Moreover, they do not include the difference introduced through even and odd $q$ states. This difference was considered in calculations of $T_c$ by Bogachek {\it et al.\/} \cite{kulik:75} in the one-channel limit. 
In Eq.~(\ref{37}) the value of $\Delta(0)$ is in fact different for even and odd $q$.
Although quantitative predictions of $T_c-T^*$ of the theory presented here might be too large compared to the experiment; it serves as an upper limit, because it describes the maximum possible persistent current. Scattering processes in real systems will further reduce $T_c-T^*$.

For temperatures close to $T_c$, the difference of the eigenenergies of even and odd $q$ states is less important than at $T=0$. Thus the deviation from the $hc/e$-periodicity of the current and of the order parameter is smaller. Furthermore, persistent currents in the normal state are exponentially small compared to the persistent supercurrents below $T_c$. Their respective $hc/e$-periodic behavior is therefore essentially invisible in the flux regime where $\Delta=0$.  For the annulus described in Fig.~\ref{Fig6}, the difference between $\Delta(\phi=0)$ and $\Delta(\phi=1/2)$ is still visible, but the corresponding differences in the current are too small.

\section{ Conclusions}\label{sec:conclusion}
We have described the crossover from the $hc/e$-periodic persistent currents as a function of magnetic flux in a metallic loop to the $hc/2e$-periodic persistent supercurrent in a $1D$ loop as well as in a multi-channel annulus. While a $1D$ superconducting ring is a rather idealized system, it proves valuable for discussing the physics of this crossover.   
A ring with a radius smaller than half the superconducting coherence length, shows an $hc/e$-periodic super current, which reaches the critical current at a critical flux value $\phi_c$,
determined by the flux dependent closing of the gap. 
Assuming that this relation remains unchanged on a ring with 
finite thickness $d\ll R$, as indeed suggested by the multi-channel model, $R_c$ would be of the order of $1\mu$m for aluminum rings.
In two or three dimensions, 
$\Delta$ remains finite beyond $\phi_c$. 
The temperature controlled crossover, while cooling through $T_c$, appears within a temperature window proportional to $1/R^2$ and thus appears hard to detect in experiment. 

\begin{center} Acknowledgements \end{center}

We are grateful to Yuri Barash, John Kirtley, Christof Schneider anf Jochen Mannhart for useful discussions.
This work was supported by the Deutsche Forschungsgemeinschaft through SFB 484,
the EC (Nanoxide), and the ESF (THIOX).

\appendix

\section{Doppler Shift and Nodes of the Hankel Function Ansatz}
The ansatz  for $u_{\bf n}(r)$ and $v_{\bf n}(r)$ with two Hankel functions  (Eqs.~(\ref{7.5}) and (\ref{7})) solves the normal state Schr\"odinger equation for the annulus as well as the BdG equations in the superconducting state with integer and half-integer flux values.
In this appendix we show that it is not possible to construct an approximate analytic solution for the superconducting annulus that includes the effect of the Doppler shift. In this case $u_{\bf n}(r)$ and $v_{\bf n}(r)$ have the independent eigenenergies $(\hbar^2\!/2m)\,{\gamma_{\bf n}^{u}}^2$ and $(\hbar^2\!/2m)\,{\gamma_{\bf n}^{v}}^2$.

For this purpose we analyze the relation between the Doppler shift of the eigenfunctions of the annulus in the normal state ($\Delta=0$) and the shift of their nodes with respect to the radial coordinate,
using the following asymptotic form for the Hankel functions \cite{abramowitz}:
\begin{align}
H^{(1,2)}_l\!\left(\frac{l}{\cos x}\right)=\sqrt{\!\frac{2}{\pi\,l \tan x}}\exp\!\left[\pm i\left(l\tan x-l\,x-\!\frac{\pi}{4}\right)\!\right]\!,
\label{12}
\end{align}
which approximates $H^{(1)/(2)}_l$ for $l\gg1$.
Choosing $x=\text{acos}(l/\gamma r)$ leads with $\tan(\arccos\,x)=\sqrt{1-x^2}/x$ to
\begin{multline}
H_l^{(1,2)}(\gamma r)=\sqrt{\frac{2}{\pi l}}\left[\left(\frac{\gamma r}{l}\right)^2-1\right]^{-1/4}\\
\times\exp\left[\pm i\left(r\sqrt{\gamma^2-\frac{l^2}{r^2}}-l\,\arccos\,\frac{l}{\gamma r}-\frac{\pi}{4}\right)\right].
\label{13}
\end{multline}
Thus Eq.~(\ref{13}) approximates $H_l^{(1,2)}(\gamma r)$ for $\gamma r\gg1$. Inserting Eq.~(\ref{13}) into the boundary conditions (\ref{9}) determines the constants $c^\alpha_{\bf n}$ and $\gamma_{\bf n}^\alpha$:
\begin{align}
\begin{split}
c^\alpha_{\bf n}=\exp\left[2i\left(D^\alpha_{\bf n}(R_1)-\frac{\pi}{4}\right)\right]
=\exp\left[2i\left(D^\alpha_{\bf n}(R_2)-\frac{\pi}{4}\right)\right],
\end{split}
\label{14}
\end{align}
with 
\begin{align}
D^\alpha_{\bf n}(r)=r\sqrt{\left(\gamma^\alpha_{\bf n}\right)^2-\frac{l_\alpha^2}{r^2}}-l_\alpha\,\arccos\,\frac{l_\alpha}{\gamma^\alpha_{\bf n}r}.
\label{16}\end{align}
The wave functions $u_{\bf n}(r)$ and $v_{\bf n}(r)$ (Eqs.~(\ref{7.5},\ref{7})) become
\begin{multline}
u_{\bf n}(r)=u_{\bf n}\sqrt{\frac{8}{\pi l_u}}\left[\left(\frac{\gamma^u_{\bf n}r}{l_u}\right)^2-1\right]^{-1/4}\\
\ e^{i\left[D^u_{\bf n}(R_1)+\frac{\pi}{4}\right]}\sin\left[D^u_{\bf n}(r)-D^u_{\bf n}(R_1)\right],
\label{17}
\end{multline}
\begin{multline}
v_{\bf n}(r)=v_{\bf n}\sqrt{\frac{8}{\pi l_v}}\left[\left(\frac{\gamma^v_{\bf n}r}{l_v}\right)^2-1\right]^{-1/4}\\
e^{i\left[D^v_{\bf n}(R_1)+\frac{\pi}{4}\right]}\sin\left[D^v_{\bf n}(r)-D^v_{\bf n}(R_1)\right].
\label{18}
\end{multline}
The vanishing of the wavefunction for $r=R_2$ therefore implies that 
\begin{align}
D^\alpha_{\bf n}(R_2)-D^\alpha_{\bf n}(R_1)=-\pi\rho
\label{40}
\end{align}
for an integer $\rho$, which determines $\gamma_{\bf n}^\alpha$. In the limit of a thin annulus ($R_1\gg R_2-R_1$), we expand $D^\alpha_{\bf n}(r)$ in $1/r$ and find
\begin{align}
D^\alpha_{\bf n}(r)-D^\alpha_{\bf n}(R_1)\approx(r-R_1)\left[\gamma^\alpha_{\bf n}-\frac{l_\alpha^2}{2\gamma^\alpha_{\bf n}rR_1}\right].
\label{20}
\end{align}
With this asymptotic form the boundary condition (\ref{40}) becomes a quadratic equation in $\gamma^\alpha_{\bf n}$:
\begin{align}
({\gamma^\alpha_{\bf n}})^2-\frac{\pi\rho}{R_1-R_2}\gamma^\alpha_{\bf n}-\frac{l_\alpha^2}{2R_1R_2}=0,
\label{21}
\end{align}
which has the positive solution
\begin{align}
\gamma^\alpha_{\bf n}=\frac{1}{2}\left[\frac{\pi\rho}{R_1-R_2}+\sqrt{\left(\frac{\pi\rho}{R_1-R_2}\right)^2+\frac{l_\alpha^2}{2R_1R_2}}\right].
\label{22}
\end{align}
This is the simplest possible approximation for the eigenenergies of the uncoupled equations ($\Delta=0$) of the annulus containing the Doppler shift, which is controlled by $l_\alpha^2$. The flux $\phi$ enters $l_u$and $l_v$ with different signs (see Eq.~\ref{5})). Thus, if $\gamma^u_{\bf n}$ decreases as a function of $\phi$, $\gamma^v_{\bf n}$ increases. Since $q-2\phi<1$ in the ground state, the Doppler shift $(\hbar^2/2m)[{\gamma^\alpha_{\bf n}}^2(q-2\phi)-{\gamma^\alpha_{\bf n}}^2(0)]$ is linear in leading order in $(q-2\phi)/\sqrt{R_1}$.

We further find the nodes $r_{{\bf n}m}$ of $u_{\bf n}(r)$ and $v_{\bf n}(r)$ by setting expression (\ref{20}) equal to $\pi m$, where $m$ is a positive integer, and solving it for $r>0$:
\begin{multline}
r_{{\bf n}m}=\frac{1}{2}\Bigg[R_1+\frac{l_\alpha^2}{2{\gamma^\alpha_{\bf n}}^2R_1}-\frac{\pi m}{\gamma^\alpha_{\bf n}}\\
+\sqrt{\left(R_1+\frac{l_\alpha^2}{2{\gamma^\alpha_{\bf n}}^2R_1}-\frac{\pi m}{\gamma^\alpha_{\bf n}}\right)^2-\frac{2l_\alpha^2}{{\gamma^\alpha_{\bf n}}^2}}\Bigg].
\label{24}
\end{multline}
The shift of the nodes $r_{{\bf n}m}(q-2\phi)-r_{{\bf n}m}(0)$ as a function of flux is again linear in $(q-2\phi)/\sqrt{R_1}$ to leading order. Thus both the Doppler shift and the nodes of $u_{\bf n}(r)$ shift linearly with $\phi$ and conversely when compared with the Doppler shift and the nodes of $v_{\bf n}(r)$. 

The coupled Eqs.~(\ref{8}) for $\Delta\neq0$ resulting from the ansatz~(\ref{7.5},\ref{7}) with non integer (or non half-integer) flux can be solved only by wave functions $u_{\bf n}(r)$ and $v_{\bf n}(r)$ with the same $r$-dependence. To obtain a solution of this problem, one can expand the wave functions as a sum of Hankel functions and numerically solve for the coefficients or directly solve the coupled differential equations numerically.



\end{document}